\documentclass[sn-mathphys-num]{sn-jnl}% Math and Physical Sciences Numbered Reference Style 

\usepackage{graphicx}%
\usepackage{multirow}%
\usepackage{amsmath,amssymb,amsfonts}%
\usepackage{amsthm}%
\usepackage{mathrsfs}%
\usepackage[title]{appendix}%
\usepackage{xcolor}%
\usepackage{textcomp}%
\usepackage{manyfoot}%
\usepackage{booktabs}%
\usepackage{algorithm}%
\usepackage{algorithmicx}%
\usepackage{algpseudocode}%
\usepackage{listings}%

\theoremstyle{thmstyleone}%
\theoremstyle{thmstyletwo}%

\theoremstyle{thmstylethree}%

\begin{document}

\title[Article Title]{FEBio FINESSE: An open-source finite element simulation approach to estimate \textit{in vivo} heart valve strains using shape enforcement }

\author[1,2]{\fnm{Devin W.} \sur{Laurence}}\email{laurenced@chop.edu}

\author[1]{\fnm{Patricia M.} \sur{Sabin}}\email{sabinp@chop.edu}

\author[1]{\fnm{Analise M.} \sur{Sulentic}}\email{sulentica@chop.edu}

\author[1]{\fnm{Matthew} \sur{Daemer}}\email{daemerm@chop.edu}

\author[3,4]{\fnm{Steve A.} \sur{Maas}}\email{steve.maas@utah.edu}

\author*[3,4]{\fnm{Jeffrey A.} \sur{Weiss}}\email{jeff.weiss@utah.edu}

\author*[1,2]{\fnm{Matthew A.} \sur{Jolley}}\email{jolleym@chop.edu}

\affil[1]{\orgdiv{Department of Anesthesiology and Critical Care Medicine}, \orgname{Children's Hospital of Philadelphia}, \orgaddress{\city{Philadelphia}, \state{PA}, \country{USA}}}

\affil[2]{\orgdiv{Division of Cardiology}, \orgname{Children's Hospital of Philadelphia}, \orgaddress{\city{Philadelphia}, \state{PA}, \country{USA}}}

\affil[3]{\orgdiv{Department of Biomedical Engineering}, \orgname{University of Utah}, \orgaddress{\city{Salt Lake City}, \state{UT}, \country{USA}}}

\affil[4]{\orgdiv{Scientific Computing Institute}, \orgname{University of Utah}, \orgaddress{\city{Salt Lake City}, \state{UT}, \country{USA}}}

%%==========%%
%% Abstract %%
%%==========%%

\abstract{Finite element simulations are an enticing tool to evaluate heart valve function in healthy and diseased patients; however, patient-specific simulations derived from 3D echocardiography are hampered by several technical challenges. In this work, we present an open-source method to enforce matching between finite element simulations and \textit{in vivo} image-derived heart valve geometry in the absence of patient-specific material properties, leaflet thickness, and chordae tendineae structures. We evaluate FEBio Finite Element Simulations with Shape Enforcement (FINESSE) using three synthetic test cases covering a wide range of model complexity. Our results suggest that FINESSE can be used to not only enforce finite element simulations to match an image-derived surface, but to also estimate the first principal leaflet strains within $\pm0.03$ strain. Key FINESSE considerations include: (i) appropriately defining the user-defined penalty, (ii) omitting the leaflet commissures to improve simulation convergence, and (iii) emulating the chordae tendineae behavior via prescribed leaflet free edge motion or a chordae emulating force. We then use FINESSE to estimate the \textit{in vivo} valve behavior and leaflet strains for three pediatric patients. In all three cases, FINESSE successfully matched the target surface with median errors similar to or less than the smallest voxel dimension. Further analysis revealed valve-specific findings, such as the tricuspid valve leaflet strains of a 2-day old patient with HLHS being larger than those of two 13-year old patients. The development of this open source pipeline will enable future studies to begin linking \textit{in vivo} leaflet mechanics with patient outcomes}


\maketitle

% \newpage

%%==============%%
%% Introduction %%
%%==============%%

\section{Introduction}\label{Sec:Introduction}

The atrioventricular heart valves (AHVs) regulate blood flow between the atria and ventricles, and valve incompetence can lead to atrioventricular valve regurgitation (AVVR). This undesired blood flow from the ventricle into the atrium results in circulatory inefficiency which is in turn associated with significant morbidity and mortality~\cite{otto2021}. The causes of AVVR can be either \textit{primary} (i.e., organically diseased tissue) or \textit{secondary} to another cardiac pathology, such as ischemic cardiomyopathy. In adults, ischemic mitral valve (MV---left-sided AHV) disease affects many patients with coronary artery disease and more than doubles the risk of cardiovascular mortality over 3 years~\cite{sannino2017, howsmon2020}. Similarly, the tricuspid valve (TV---right-sided AHV) is increasingly being recognized for contributions to cardiac dysfunction, with new focus on studying the effects of failure as well as novel minimally invasive therapies to maximize function~\cite{hahn2016, hahn2022, lurz2021, nickenig2019, von2023}. For children and young adults, heterogeneous populations of patients with congenital heart disease (CHD) suffer from AHV dysfunction \cite{ho2020, atz2011, kaza2011, gellis2023}. In particular, patients with palliated single ventricle physiology are disproportionately affected by AVVR, which is associated with a nearly 2.5 fold increased risk of death or transplant \cite{king2019}. Understanding the etiology, progression, and treatment of AVVR is paramount to improve outcomes for these patients.

Extensive progress has been made in the evaluation of adult and pediatric AHV function over the last two decades~\cite{myerson2012, lasso2022, guerrero2020}. Multi-modal cardiac imaging (e.g., computed tomography---CT, 3D echocardiography---3DE, or magnetic resonance imaging---MRI) now allows complete 3D reconstruction of the heart valves throughout the cardiac cycle (i.e. 4-dimensional reconstruction(4D)). 4D valve images have allowed investigation of dynamic valve structure to function~\cite{salgo2002, levack2012, lee2013} and further understanding of  the connection between cardiac loads and  cellular responses associated with valve dysplasia and functional failure~\cite{kodigepalli2020, kruithof2020, markby2020}. For example, multiple studies have now demonstrated that increased leaflet strain is associated with pathological leaflet changes, moderate or greater AVVR, and clinical valve repair failure in adults with MV disease~\cite{el2021a, el2021b, marsan2021, narang2021}. These findings have been corroborated by large animal studies, where increased leaflet tissue strain results in cell-driven pathological adaptation of the valve leaflets~\cite{meador2020}. Thus, there is an increasing desire to develop and apply methods to better understand the interplay between valve function, tissue mechanics, and cellular biosynthesis and the development of valve pathology associated with valve failure\cite{howsmon2020, nam2022, nam2023, rego2022vivo, salgo2002}. This study specifically seeks to build an open-source tool to estimate \textit{in vivo} leaflet strains, which can be further related to valve pathology or failure. 

Finite element (FE) simulations can be used to predict AHV function and understand the underlying tissue biomechanics (e.g. strain)~\cite{sacks2019}. Towards this end, investigators have developed in-house solvers (e.g.,~\cite{johnson2021}), leveraged commercial software packages (e.g., Abaqus~\cite{lee2017,sacks2019,laurence2020}), or contributed to open-source frameworks (e.g.,~\cite{wu2022, wu2023}). Initial work was largely focused on building FE simulations as a proof-of-concept~\cite{kunzelman1993, stevanella2010}. More recent studies have leveraged FE simulations to investigate relationships between valve structure, function, and mechanics. This collective effort has paved the way for future application of pre-operative simulations to inform patient-specific valve repair~\cite{lee2017, laurence2020, haese2024, mathur2024}. However, traditional use of these tools for individualized predictions of AHV function are hampered by several challenges: (i) estimating material properties~\cite{ross2024}, (ii) acquiring chordae tendineae structure geometry~\cite{khalighi2017}, and (iii) considering realistic heterogeneous valve thickness~\cite{lin2023}. Collectively, these increase the likelihood that traditional forward FE simulations may not accurately predict patient-specific AHV function.

Two classes of methods have emerged to improve FE predictions of \textit{in vivo} AHV function and thereby our understanding of the leaflet biomechanics (e.g., strain). The first, more traditional, approach is commonly known as inverse FE analysis~\cite{lee2014, ross2024}. This iterative approach optimizes some model input, such as the constitutive model parameters, so the FE prediction matches the  3D image-derived valve leaflet surface. In principle, this method can produce FE simulations with excellent standalone predictions of the AHV function, and the optimized parameters can be evaluated to non-invasively understand biomechanical changes to tissue in disease. However, this method has several practical shortcomings: (i) inverse FE problems can be ill-posed for AHVs due to poor convexity and smoothness of the search space~\cite{ross2024}, leading to non-unique or poor solutions and (ii) the FE simulation is often computationally expensive, resulting in intractable computation times when combined with the iterative nature of optimization. These challenges have led to more focus on a second approach where additional constraints are included to enforce \textit{in-silico} predictions to agree with the AHV deformation observed \textit{in vivo} . Historically, this has been achieved with elastic regularization via image registration~\cite{phatak2007, phatak2009}, whereas more recent efforts (including the present work) include corrective loads within the FE simulation to enforce the FE predictions to match the \textit{in vivo} AHV configuration~\cite{rego2018, narang2021}.

The objective of this work is to establish an open-source workflow to enforce agreement between FE simulations of AHV function and routine clinical 3DE to facilitate assessment of the \textit{in vivo} tissue biomechanics (e.g., strain). We establish a two-step approach in which the target valve configuration is estimated using our established FEBio FE simulation framework~\cite{wu2022, wu2023} and then shape enforced via a surface-to-surface registration algorithm~\cite{ateshian2010}. This method is evaluated using three synthetic test cases with varying degrees of complexity that emulate characteristics of AHV function. Key parameters are varied to understand how they influence the shape enforcement results on the basis of matching the target surface and estimating the tissue strains. The optimal shape enforcement approach is finally used to estimate the \textit{in vivo} heart valve strains for three pediatric patients with varying cardiac physiology, such as complex congenital heart disease. 

%%=========%%
%% Methods %%
%%=========%%

\section{Methods}\label{Sec:Methods}

We developed an open-source workflow to force forward FE simulations to match 3DE images and estimate \textit{in vivo} AHV leaflet strains. We assumed the valve geometry was available in two configurations: the undeformed reference configuration and the deformed target configuration. The target configuration could be ideally predicted via forward FE simulations (see Section~\ref{Sec:MethodsForwardFE}), but this is not yet feasible in general for patient-specific cases due to uncertanty in : (i) chordae tendineae structure geometry, (ii) precise loading conditions (i.e., transvalvular pressure), and (iii) patient-specific material properties. Therefore, we made \textit{a priori} assumptions for these model inputs, and implemented an approach to match the target surface via shape enforcement (see Section~\ref{Sec:MethodsShapeEnforcement}) and thereby estimate \textit{in vivo} strains. Our two-step approach for \underline{Fin}ite \underline{E}lement \underline{S}imulations with \underline{S}hape \underline{E}nforcement (``FINESSE") was evaluated using three synthetic test cases. This approach was then used to estimate \textit{in vivo} leaflet strains for three pediatric patients (two with complex CHD). All models used in the analyses for this project can be accessed in the FEBio Model Repository (Section 3.10 of \cite{FEBioUser} or \url{https://repo.febio.org/modelRepo/}). 

\subsection{\underline{Fin}ite \underline{E}lement \underline{S}imulations with \underline{S}hape \underline{E}nforcement (FINESSE)} \label{Sec:MethodsFE}

We refer to three configurations: (i) the undeformed valve configuration ($\Omega_{0}$ -- mid-diastole), (ii) the deformed valve configuration ($\Omega_{\textrm{Target}}$ -- mid-systole), and (iii) the FE prediction at any time $t$ during FINESSE ($\Omega_{\textrm{FE}}(t)$). All three configurations ($\Omega_{0}$, $\Omega_{\textrm{Target}}$, and $\Omega_{\textrm{FE}}$) have associated leaflet surfaces ($\mathcal{S}_{0}$, $\mathcal{S}_{\textrm{Target}}$, and $\mathcal{S}_{\textrm{FE}}$) discretized using shell elements (e.g., triangular or quadrilateral) with associated nodes ($p_{0}$, $p_{\textrm{Target}}$, and $p_{\textrm{FE}}$) and 3D locations ($\mathbf{x}_{0}$, $\mathbf{x}_{\textrm{Target}}$, and $\mathbf{x}_{\textrm{FE}}$). The elements are defined such that the surface normal ($\mathbf{n}$) points towards the atrium for the FE simulation ($\mathcal{S}_{\textrm{FE}}(t)$) and the ventricle for the known surfaces ($\mathcal{S}_{0}$ and $\mathcal{S}_{\textrm{Target}}$). This will ensure the surface-to-surface registration algorithm penalizes separation rather than proximity; however, the direction of the surface normal vectors can be changed as part of the \texttt{sliding-elastic} interface settings in FEBio.

\subsubsection{FINESSE Step 1: Forward Finite Element Simulation}
\label{Sec:MethodsForwardFE}
The target configuration ($\Omega_{\textrm{Target}}$) was first estimated using our FE simulation approach with FEBio \cite{wu2022, wu2023}. The heart valve leaflet surface ($\mathcal{S}_{0}$) was discretized with linear quadrilateral shell elements with an assumed thickness of $0.39$\,mm \cite{wu2023}. The nonlinear nearly-incompressible mechanical behavior was modeled using the hyperelastic Lee-Sacks constitutive model with strain energy density ($\psi$)

\begin{equation}
\label{Eq:LeeSacks}
    \psi = \frac{c_0}{2}\left(I_{1}-3\right) + \frac{c_1}{2}\left(e^{c_{2}\left(I_{1}-3\right)^2}-1\right).
\end{equation}

\noindent Here, $c_0$, $c_1$, $c_2$ are material coefficients, $I_1$ is the first invariant of the right Cauchy deformation tensor $\mathbf{C}=\mathbf{F}^T\mathbf{F}$, and $\mathbf{F}$ is the deformation gradient. While it is well-known that heart valve leaflets have anisotropic mechanical behaviors stemming from aligned collagen fiber architectures~\cite{jett2020, fitzpatrick2022}, the prior work of Wu \textit{et al}.~\cite{wu2018} suggested that isotropic constitutive models are sufficient to capture heart valve closure in numerical simulations. The chordae tendineae were represented by tension-only springs that generated $2000$\,mN of force between stretches $1.01$ and $1.03$~\cite{ross2020}. The simplified chordae tendineae were only considered to obtain the synthetic target solutions for numerical validation of FINESSE (see Section~\ref{Sec:MethodsSynthetic}). Otherwise, a chordae emulating force (CEF) was prescribed to the leaflet free edge nodes in the direction normal to the annulus~\cite{rego2018}. 

Valve closure was simulated by prescribing a linearly ramped pressure over $0.05$\,s to the ventricular leaflet surface. Case-specific pressure magnitudes are defined in Section~\ref{Sec:MethodsSynthetic}. We assumed the annulus was pinned for the three synthetic test cases, whereas the patient-specific dynamic annulus motion was prescribed for the three \textit{in vivo} cases. Dynamic analysis was performed using the default FEBio automatic time stepping algorithm with maximum time step of $10^{-4}$\,s and minimum time step of $10^{-8}$\,s. All simulations were completed using available HPC resources at the Children's Hospital of Philadelphia. 

\subsubsection{FINESSE Step 2: Shape Enforcement}
\label{Sec:MethodsShapeEnforcement}
The target leaflet surface mesh ($\mathcal{S}_{\textrm{Target}}$) was defined as a fixed rigid body. This surface was the leaflet mesh in the final FE simulation time step for the synthetic test cases (Section~\ref{Sec:MethodsSynthetic}) and the patient-specific segmentation for the \textit{in vivo} cases (Section~\ref{Sec:MethodsInVivo} and Appendix~\ref{Sec:AppendixB}). The predicted valve configuration from the forward FE simulation in Section~\ref{Sec:MethodsForwardFE} ($\Omega_{\textrm{FE}}(t)$) was then enforced to match the target surface ($\mathcal{S}_{\textrm{Target}}$) using a \texttt{sliding-elastic} interface with the tension flag enabled~\cite{ateshian2010}. The gap function ($g$) was defined as the surface-surface distance projected onto the local FE surface normal ($\mathbf{n}_{\textrm{FE}}$)

\begin{equation}
\label{Eq:GapStep1}
    \mathbf{x}_{\textrm{Target}, j} = \mathbf{x}_{\textrm{FE}, i} + g_{i}\mathbf{n}_{\textrm{FE}, i},
\end{equation}

\noindent where, $\mathbf{x}_{\textrm{Target}, j}$ is the intersection between the target surface and a ray emanating from ($\mathbf{x}_{\textrm{FE}, i}$) in the direction $\mathbf{n}_{\textrm{FE}, i}$. This is rearranged to obtain a direct relationship for $g$

\begin{equation}
    g_{i} = (\mathbf{x}_{\textrm{Target}} - \mathbf{x}_{\textrm{FE}, i})\cdot\mathbf{n}_{\textrm{FE}, i}.
\end{equation}

\noindent The contact gap $g_i$ was minimized by applying a contact traction normal to leaflet surface with magnitude determined via the standard penalty method, i.e., $t_{n, i} = \epsilon g_i$. Note $\epsilon$ is a penalty factor with larger values resulting in smaller contact gap $g$; however, determining an appropriate value of $\epsilon$ can depend on the material coefficients and other geometric factors (e.g., thickness). Therefore, FEBio allows the user to instead define a \textit{penalty scale factor}, which is then combined with the material and geometrical properties to determine $\epsilon$ (Section 7.1.7 of \cite{FEBioTheory}). Nevertheless, large values of $\epsilon$ may worsen simulation convergence~\cite{ateshian2010}, so we used a subsequent Augmented-Lagrangian step \cite{simo1992} to further minimize $g$ while limiting the value of $\epsilon$.

\begin{figure}[tbp]
\includegraphics[width=0.75\textwidth]{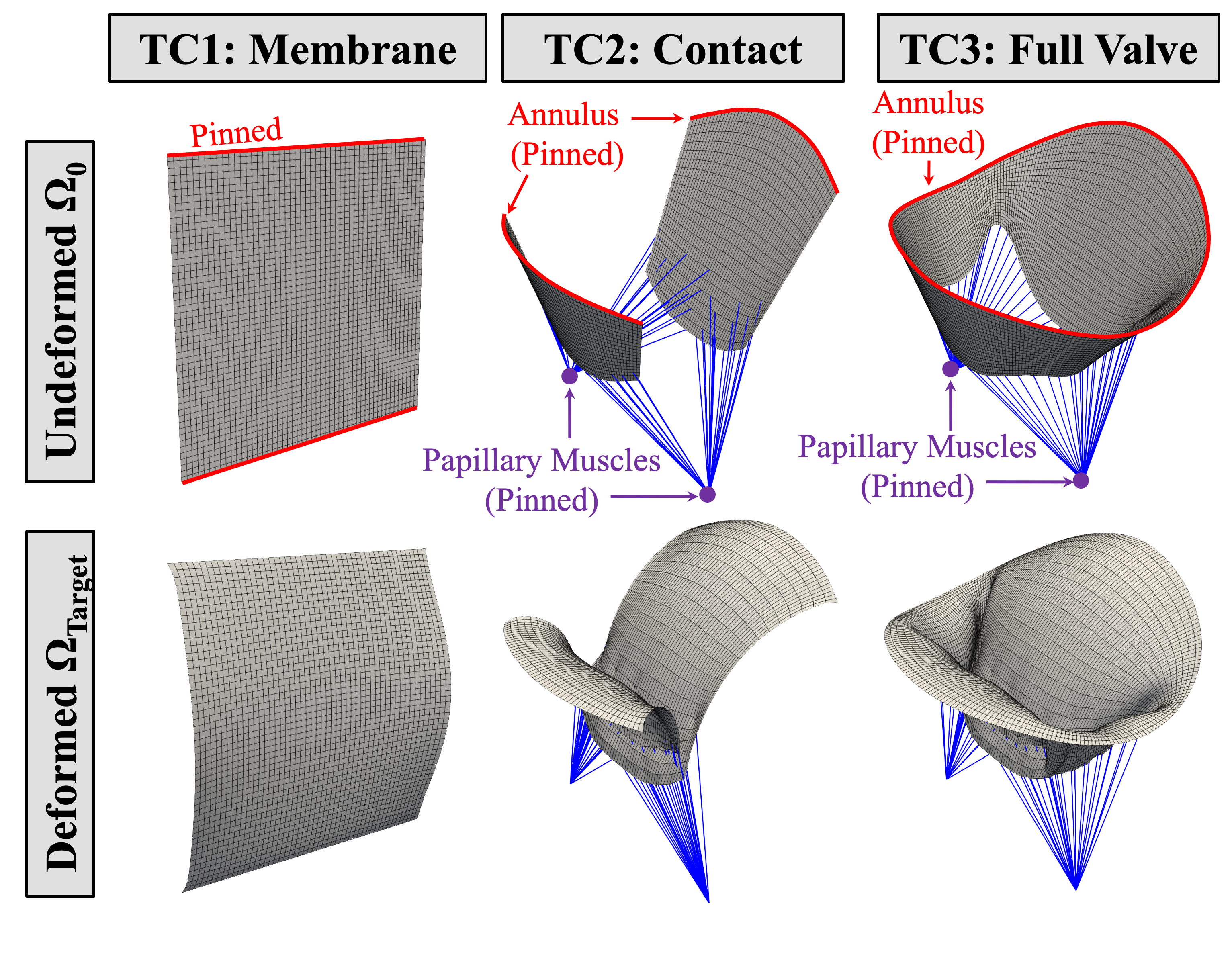}
\centering
\caption{Synthetic FE simulations to verify FINESSE predictions. The (\textit{top}) initial geometries underwent pressurization (see Section~\ref{Sec:MethodsSynthetic} to achieve the (\textit{bottom}) deformed configurations, which are used as the target surfaces in FINESSE. (TC: Test Case)}
\label{Figure:SyntheticDefinition}
\end{figure}

\subsection{Synthetic Test Cases}\label{Sec:MethodsSynthetic}
Three synthetic test cases were used to evaluate FINESSE performance. Each test case was used for three types of numerical investigations (see Section~\ref{Sec:MethodsSyntheticInvestigations}) to understand how FE simulation parameters and FINESSE parameters may influence performance. For Test Case 2 and Test Case 3, chordae tendineae were considered to generate the synthetic FE simulation data, whereas a CEF was applied during FINESSE in agreement with our \textit{in vivo} pipeline.

\subsubsection{Test Case 1: Membrane}
\textbf{Test Case 1} (Figure~\ref{Figure:SyntheticDefinition}(a)) considered the pressurization of a thin membrane---a basic component of valve closure. The outer boundary was pinned and a $13.3$\,kPa pressure ($100$\,mmHg) was applied to one surface of the membrane. 

\subsubsection{Test Case 2: Simplified Leaflet Contact}
\textbf{Test Case 2} (Figure~\ref{Figure:SyntheticDefinition}(b)) considered the contact between two leaflets to avoid challenges associated with folding in the commissures. A transvalvular pressure of $13.3$\,KPa ($100$\,mmHg) was applied to the ventricular leaflet surface. 

\subsubsection{Test Case 3: Stereotypical Mitral Valve}
\textbf{Test Case 3} (Figure~\ref{Figure:SyntheticDefinition}(c)) considered the closure of a stereotypical MV geometry to fully encapsulate the complexities of modeling AHV function (i.e., self-contact, commissure folding, large deformations). A transvalvular pressure of $13.3$\,kPa ($100$\,mmHg) was applied to the ventricular leaflet surface. 

\subsubsection{Numerical Investigations}
\label{Sec:MethodsSyntheticInvestigations}
Three types of numerical investigations were performed for each test case. The \textit{first numerical investigation} aimed to understand how FE simulation parameters influenced FINESSE performance. We uniformly sampled the penalty $\epsilon$ ($[0-5]$\,kPa), the Augmented Lagrangian tolerance ($[10^{-3}-10^{-1}]$), the applied pressure ($[0-20]$\,kPa), and noise added to the target surface. The \textit{second numerical investigation} sought to understand how tissue-related parameters influenced FINESSE performance. Here, we uniformly sampled the Lee-Sacks constitutive model (Eq.~\ref{Eq:LeeSacks}) parameters ($c_0\in[10-250]$\,kPa,  $c_1\in[600-4000]$\,kPa, and $c_2\in[0.1-0.5]$) and the tissue thickness ($[0.1-0.6]$\,mm). The \textit{third numerical investigation} concerned cardiac imaging-related limitations, and focused on how to best overcome challenges segmenting the chordae tendineae and correctly identifying the leaflet commissures and coaptation region.

\subsubsection{Evaluation Criteria}
\label{Sec:MethodsSyntheticEvaluation}
FINESSE results for the synthetic cases were evaluated using two criteria: (i) fit to the target surface and (ii) estimation of leaflet strains. We used the bidirectional local distance metric (BLDM) to determine the point-wise distance ($d_{\textrm{BLDM},i}$) between $\mathcal{S}_{\textrm{Target}}$ and $\mathcal{S}_{\textrm{FE}}$ (Appendix~\ref{Sec:AppendixA}). On the other hand, we compared the first principal strains from FINESSE to values from the synthetic FE simulations to evaluate how well FINESSE recaptures tissue strains. The distributions for the BLDM and first principal strains are used where possible. Otherwise, results are presented as median [IQR].

\subsection{\textit{In Vivo} Application}\label{Sec:MethodsInVivo}
To demonstrate the use of FINESSE for \textit{in vivo} data, we estimated the \textit{in vivo} leaflet strains for three pediatric patients: (i) a healthy mitral valve from a 13-year old patient, (ii) a complete atrioventricular canal (CAVC) from a 13-year old patient, and (iii) a tricuspid valve from a 2-day old patient with hypoplastic left heart syndrome (HLHS). The use of 3DE images for these patients was approved by the Institutional Review Board at the Children's Hospital of Philadelphia. 

\subsubsection{FE Model Preparation}
A complete description of our 3DE image to FE model pipeline is provided in Appendix~\ref{Sec:AppendixB}. In brief, we used SlicerHeart modules~\cite{lasso2022} in the open-source Slicer software~\cite{fedorov2012} to generate a quadrilateral shell mesh of the leaflet geometry in the mid-diastolic frame (i.e., $\Omega_{\textrm{0}}$) and a triangular shell mesh of the leaflet atrial surface in the mid-systolic frame (i.e., $\Omega_{\textrm{Target}}$). The FE mesh nodes associated with the annulus were projected onto user-defined annular curves in mid-diastole. This projection, the valve commissure locations, and the mid-systolic annular curve were used to quantify and model the patient-specific annular motion in FINESSE. 

\subsubsection{FINESSE}
The resulting FE mesh, mid-systolic (target) surface, and annular displacement were used with FINESSE to estimate the \textit{in vivo} leaflet strains. A transvalvular pressure of $10$\,kPa ($75$\,mm Hg), chosen to provide good initial FE prediction of the target surface prior to FINESSE, was prescribed to the ventricular leaflet surface to simulate valve closure in the initial forward FE simulation. Concurrently, a CEF was prescribed to the leaflet free edge to emulate the contribution of chordae tendineae. In the subsequent shape enforcement step, only the penalty-based minimization of the gap function ($g$) was performed due to convergence issues with the additional Augmented-Lagrangian step. 

\subsubsection{Analysis}
The quality of fit provided by FINESSE was evaluated using the BLDM (see also Appendix~\ref{Sec:AppendixA}). The first principal strains were compared for the entire valve and the leaflet belly regions between all three patients. 

%%=========%%
%% Results %%
%%=========%%

\section{Results}\label{Sec:Results}

\subsection{Synthetic Validation of Shape Enforcement Method}

\subsubsection{Numerical Investigation 1: FE Simulation Parameters}
\label{Sec:FINESSEInput}
\begin{figure}[h]
\includegraphics[width=0.90\textwidth]{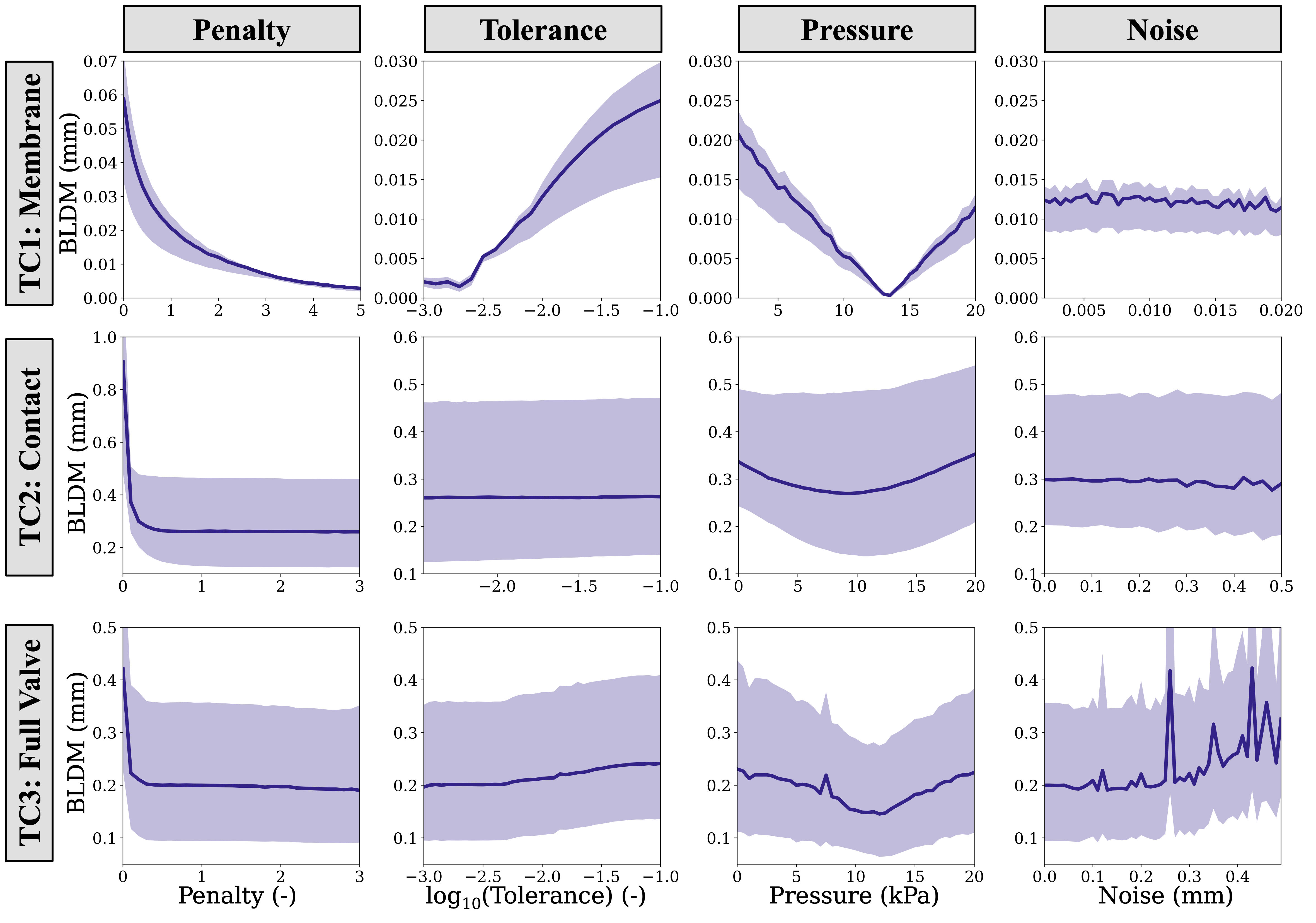}
\centering
\caption{Influence of penalty, tolerance, pressure, and noise on FINESSE performance, with more apparent trends for Test Case 1. FINESSE generally performs better with larger penalty parameters and closer proximity to the target surface (i.e., correct initial pressure). Note: solid line represents the BLDM median whereas the shaded region encompasses the BLDM IQR. (TC: Test Case, BLDM: bidirectional local distance measure)}
\label{Figure:SyntheticResults1}
\end{figure}

Trends were most apparent in Synthetic Test Case 1 (Figure\,\ref{Figure:SyntheticResults1}(a)) owing to the simplicity of modeling a pressurized membrane. In brief: (i) increasing the penalty improved BLDM ($0.003$\,$[0.001-0.003]$\,mm vs. $0.059$\,$[0.034-0.073]$\,mm), (ii) smaller Augmented Lagrangian tolerance improved BLDM ($0.002$\,$[0.001-0.003]$\,mm vs. $0.025$\,$[0.015-0.030]$), (iii) pressure had a non-monotonic relationship with applied pressures close to the synthetic pressure providing better BLDM ($0.0003$\,$[0.0002-0.0004]$\,mm vs. $0.021$\,$[0.014-0.024]$\,mm), and (iv) noise did not have a clear influence on the BLDM. Similar trends were noted for Synthetic Test Case 2 (Figure\,\ref{Figure:SyntheticResults1}(b)) and Synthetic Test Case 3 (Figure\,\ref{Figure:SyntheticResults1}(c)) although to a lesser degree. For the more complex test cases (Test Case 2 and Test Case 3), appropriately choosing the penalty value and having a good initial forward FE simulation were the most important factors. 

\subsubsection{Numerical Investigation 2: Leaflet Material Parameters}
\label{Sec:MaterialInput}
\begin{figure}[h]
\includegraphics[width=0.95\textwidth]{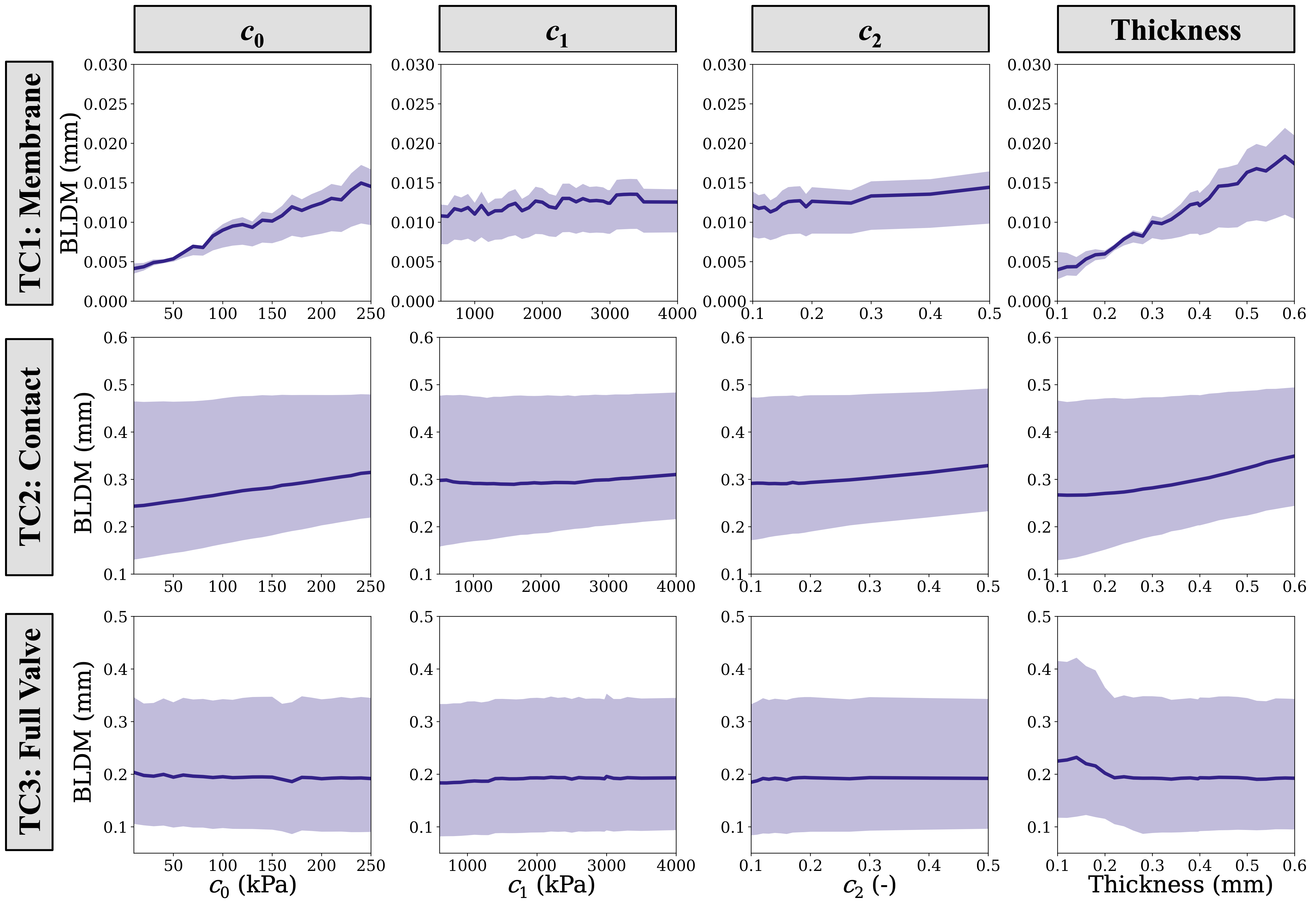}
\centering
\caption{Influence of leaflet constitutive model parameters and leaflet thickness on FINESSE performance. Softer materials (smaller values of $c_0$, $c_1$, $c_2$, and thickness) provide better agreement with the target surface (smaller BLDM).  Note: solid line represents the BLDM median whereas the shaded region encompasses the BLDM IQR. (TC: Test Case, BLDM: bidirectional local distance measure)}
\label{Figure:SyntheticResults2}
\end{figure}

As in Section\,\ref{Sec:FINESSEInput}, Test Case 1 provided the clearest illustration of the observed relationships (Figure\,\ref{Figure:SyntheticResults2}(a)). In general, smaller values of $c_0$ ($0.004$\,$[0.003-0.005]$\,mm vs. $0.015$\,$[0.010-0.017]$\,mm), $c_1$ ($0.011$\,$[0.007-0.012]$\,mm vs. $0.013$\,$[0.009-0.014]$\,mm), $c_2$ ($0.012$\,$[0.008-0.014]$\,mm vs. $0.014$\,$[0.010-0.016]$\,mm), and leaflet thicknesses ($0.004$\,$[0.003-0.006]$\,mm vs. $0.017$\,$[0.010-0.021]$\,mm) resulted in smaller median BLDM (i.e., better agreement). This suggests that FINESSE performs better with softer materials, likely due to FINESSE more easily penalizing the FE simulation to match the target configuration. These trends were shared for Test Case 2 and Test Case 3, except smaller values of tissue thickness did not improve the BLDM for Test Case 3 (Figure\,\ref{Figure:SyntheticResults2}(c)). This likely stems from thinner leaflets having different folding behaviors in the commissures. 

\subsubsection{Numerical Investigation 3: Cardiac Imaging-Related Challenges}
\label{sec:ResultsCardiacImagingChallenges}
\begin{figure}[h]
\includegraphics[width=0.95\textwidth]{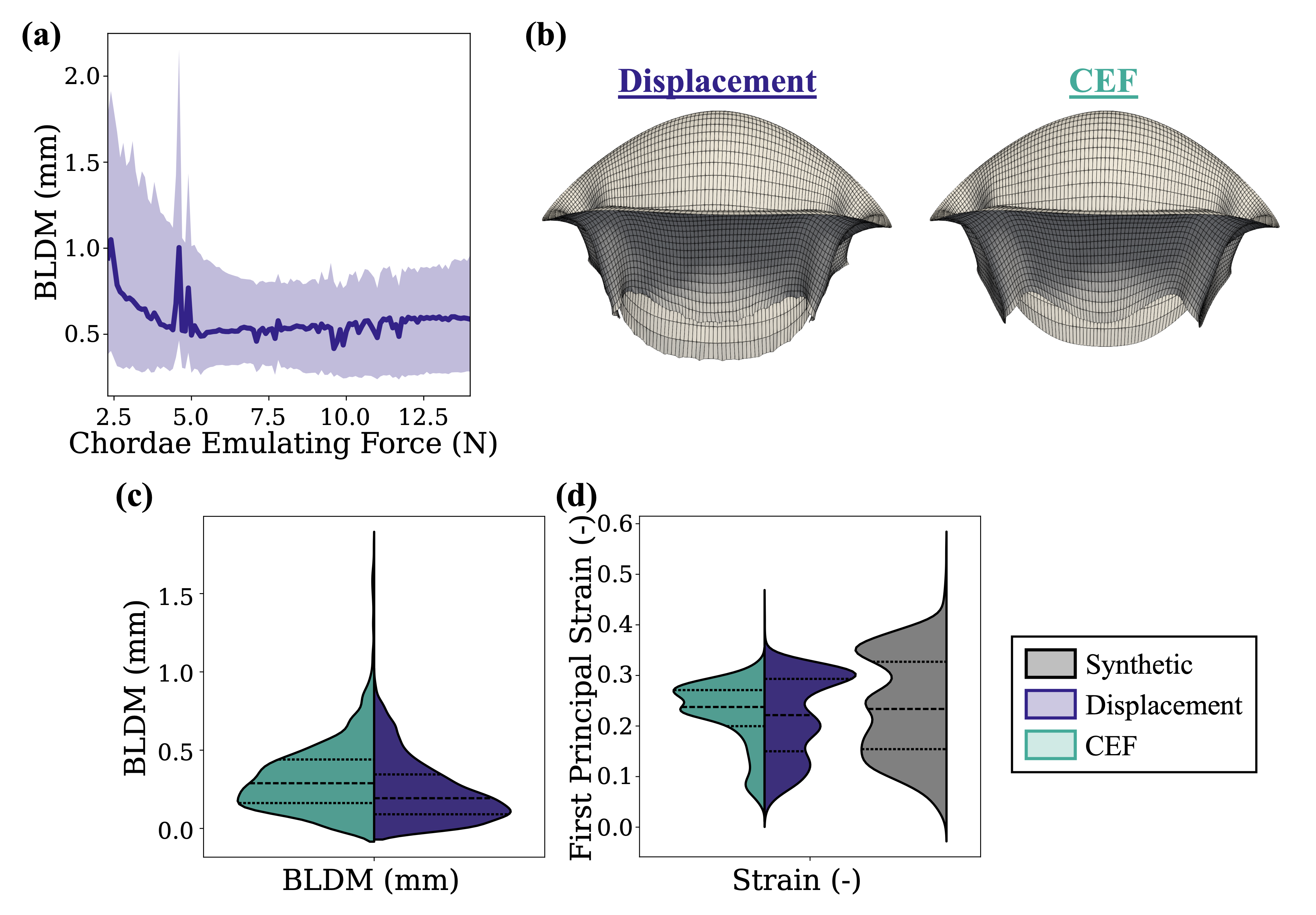}
\centering
\caption{Methods for emulating chordae function in the absence of image-derived chordae structures. (a) Relationship between the prescribed chordae emulating force (CEF) and BLDM (i.e., FINESSE performance).  Note: solid line represents the BLDM median whereas the shaded region encompasses the BLDM IQR. (b) Visual comparison of the two assumptions: (\textit{left}) displacing the free edge to the final configuration and (\textit{right}) prescribing a CEF to the leaflet free edge that is normal to the annulus plane. (c) FINESSE ability to match target surface and (d) estimation of known synthetic first principal strains.}
\label{Figure:FreeEdgeAssumptions}
\end{figure}

The motion of the leaflet free edge could ideally be extracted from the 3DE and prescribed during the FE simulation to overcome the inability to acquire chordae structures (e.g., \cite{ross2024}). However, this heavily relies on the image quality and clarity of the free edge in both systole and diastole. We therefore compared two options to overcome the lack of chordae structures from cardiac images: (i) prescribing leaflet free edge motion and (ii) prescribing a CEF to the leaflet free edge to compensate for the lack of chordae. We found the BLDM converges to approximately $0.52$\,$[0.32-0.87]$\,mm with a CEF of $6.0$\,N and a good representation of the free edge behavior (Figure\,\ref{Figure:FreeEdgeAssumptions}(b)). Furthermore, the BLDM was slightly larger for the CEF case ($0.289$\,$[0.162-0.440]$\,mm) compared to the prescribed free edge displacement ($0.193$\,$[0.090-0.345]$\,mm) (Figure\,\ref{Figure:FreeEdgeAssumptions}(c)), yet both errors were similar to the smallest approximate voxel size for \textit{in vivo} applications ($>0.2$\,mm). However, we did note that the CEF provided a more homogeneous representation of the leaflet strains compared to the prescribed free edge motion, although the median strains ($0.237$\,$[0.199-0.271]$) are similar to the prescribed displacement ($0.222$\,$[0.150-0.293]$)  and synthetic ($0.234$\,$[0.154-0.327]$) results. These results collectively demonstrate that a CEF can be used in the absence of the free edge motion, but it will lead to slightly worse agreement with the target surface and a more homogeneous distribution, yet similar central tendency, of the leaflet strains. 

\begin{figure}[h]
\includegraphics[width=0.95\textwidth]{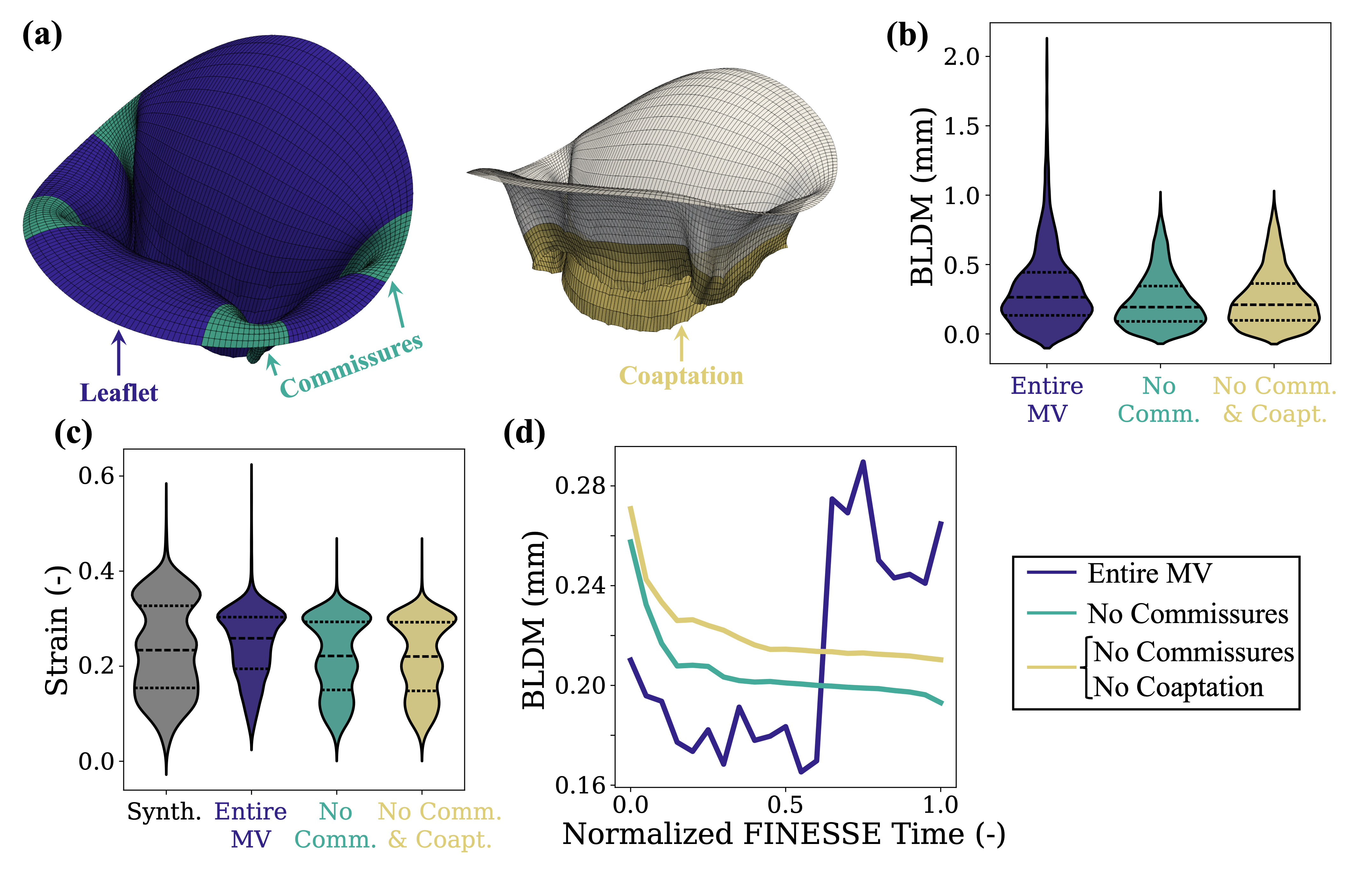}
\centering
\caption{Effect of omitting valve regions on FINESSE performance. (a) ROI definition for the commissures and coaptation region. (b) FINESSE ability to match the target surface when omitting valve regions. (c) Estimated first principal strains from FINESSE when omitting valve regions compared to known synthetic results. (e) Time evolution of BLDM throughout the FINESSE simulation highlighting the instability of using the entire leaflet geometry, thus favoring omitting at least valve commissures from FINESSE.}
\label{Figure:ValveRegions}
\end{figure}

It can be challenging to correctly segment and identify leaflets in the commissure and coaptation regions of heart valves. Therefore, we considered three scenarios: (i) using the entire MV geometry, (ii) removing the commissures, and (iii) removing the commissures and coaptation region (regions defined in Figure\,\ref{Figure:ValveRegions}(a)). All three scenarios considered the BLDM of the entire MV surface. Interestingly, we found that using the entire MV geometry in FINESSE led to the largest BLDM ($0.265$\,$[0.134-0.444]$\,mm) (Figure\,\ref{Figure:ValveRegions}(b)) and most different first principal strains ($0.259$\,$[0.194-0.303]$) compared to the synthetic case ($0.234$\,$[0.154-0.327]$) (Figure\,\ref{Figure:ValveRegions}(c)). Differences when removing the commissures compared to removing the commissures and coaptation were small for BLDM ($0.193$\,$[0.090-0.345]$\,mm vs. $0.210$\,$[0.098-0.363]$\,mm) and first principal strain ($0.222$\,$[0.150-0.293]$ vs. $0.220$\,$[0.148-0.292]$). We also noticed that the BLDM tended not to converge during FINESSE (Figure\,\ref{Figure:ValveRegions}(d)). Therefore, it is favorable to at least remove the commissures from the shape enforcement step to improve FINESSE convergence and simulation stability.

\subsubsection{Optimal Results}
\label{Sec:OptimalResults}
\begin{figure}[h]
\includegraphics[width=0.95\textwidth]{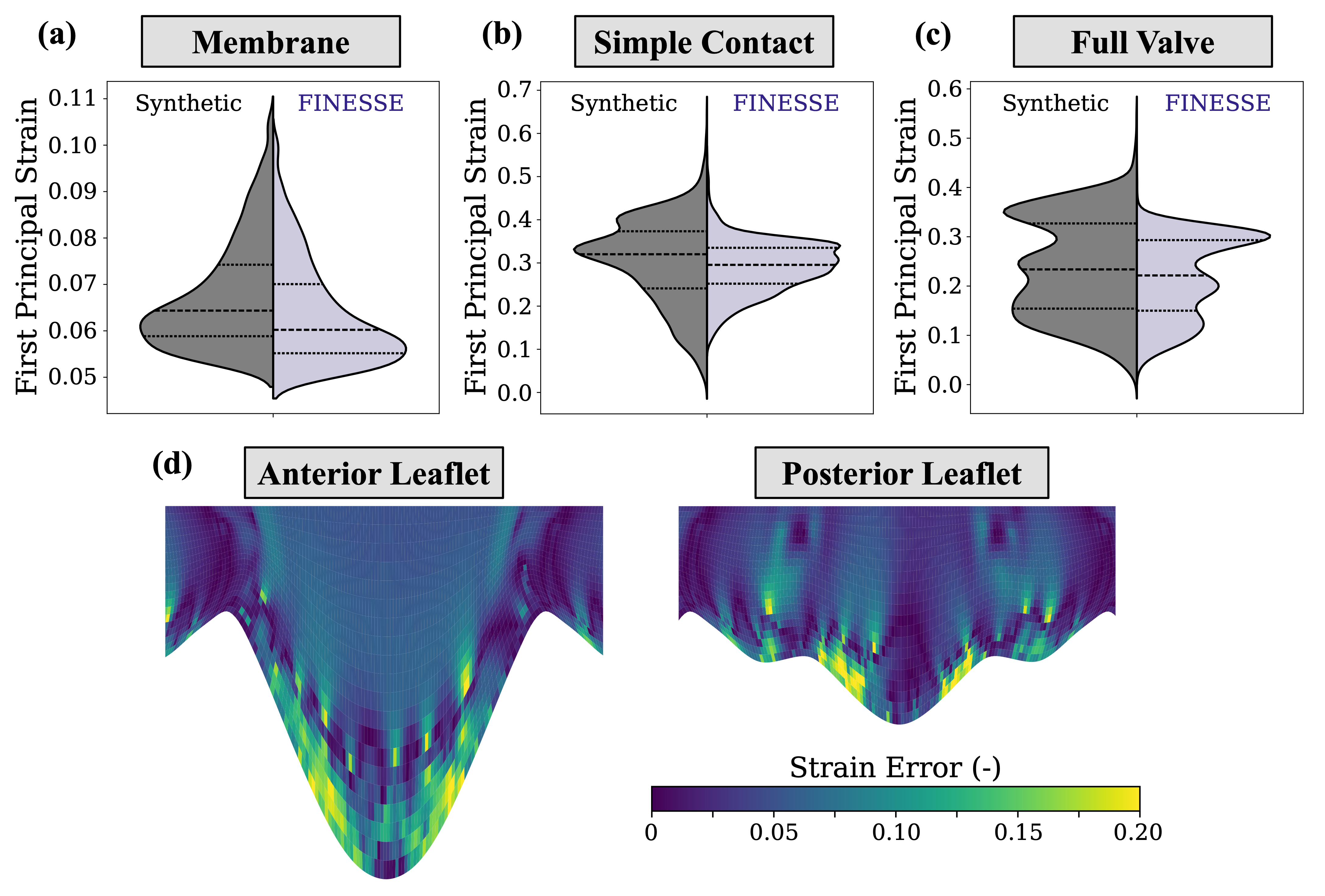}
\centering
\caption{Ability of FINESSE to recapture leaflet strains for three synthetic test cases.  Visualization of strain error on the unwrapped MV to observe regions with larger error (primarily coaptation region and commissures).}
\label{Figure:OptimalSynthetic}
\end{figure}

The estimated strains for the optimal FINESSE results of all three test cases are provided in Figure\,\ref{Figure:OptimalSynthetic}(a)-(c). FINESSE slightly under-predicted the median strain for Test Case 1 ($0.060$\,$[0.055-0.070]$ vs. $0.064$\,$[0.059-0.074]$), Test Case 2 ($0.296$\,$[0.252-0.335]$ vs. $0.320$\,$[0.241-0.373]$), and Test Case 3 ($0.222$\,$[0.150-0.293]$ vs. $0.234$\,$[0.154-0.327]$). This was likely due to the slight discrepancy between the FE simulation and target surfaces. Nevertheless, the FINESSE strain distributions resembled the known synthetic strain distributions. By further evaluating the distribution of strain error across the leaflet surface for Test Case 3 (Figure\,\ref{Figure:OptimalSynthetic}(d)), much of the discrepancy stems from the coaptation region and local large deformations stemming form chordal insertions in the synthetic simulations. 

\subsection{Estimation of \textit{In Vivo} Valve Behavior and Leaflet Strains}
\label{Sec:InVivoResults}
FINESSE provided excellent agreement with the image-derived leaflet surfaces for all three patients. Median BLDM was consistently smaller in the leaflet belly regions (MV: $0.142-0.214$\,mm, CAVC: $0.204-0.269$\,mm, HLHS: $0.127-0.284$\,mm) compared to the entire valve (MV: $0.224$\,$[0.138-0.382]$\,mm, CAVC: $0.294$\,$[0.196-0.518]$\,mm, HLHS: $0.314$\,$[0.179-0.534]$\,mm). These values are on a similar order or better than the smallest voxel dimension for each patient (MV: $0.339$\,mm, CAVC: $0.889$\,mm, HLHS: $0.277$\,mm). We further noted that the larger discrepancies (BLDM $>0.6$\,mm) were often found in the commissure regions, which was unsurprising due to their omission from the shape enforcement (see also Figure\,\ref{Figure:ValveRegions}). Nevertheless, excellent agreement in the leaflet belly regions supports the use of FINESSE for matching \textit{in vivo} heart valve behavior, and future enhancements (Section~\ref{sec:FutureExtensions}) will seek to overcome challenges with the commissures.

\begin{figure}[h]
\includegraphics[width=0.95\textwidth]{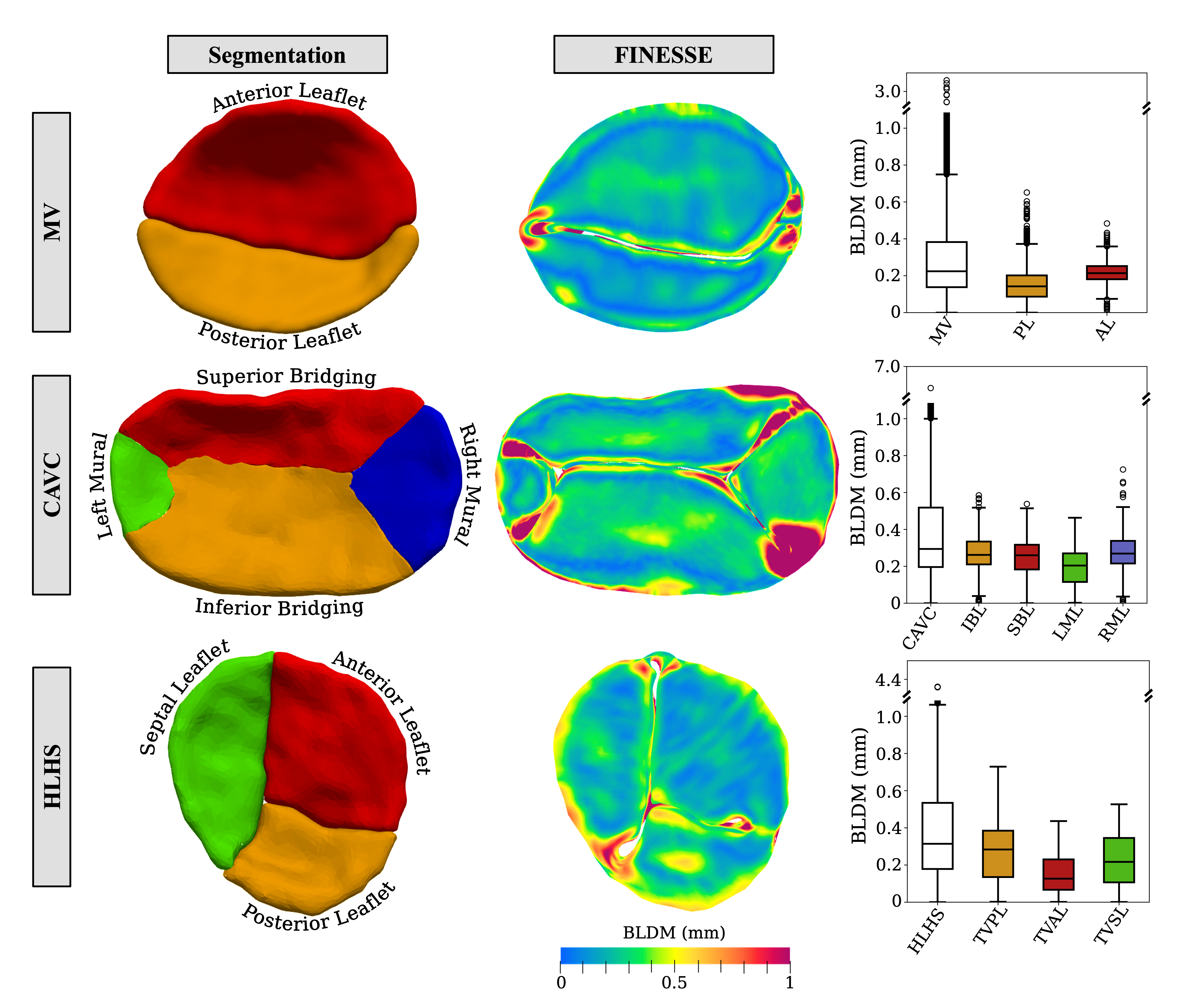}
\centering
\caption{FINESSE performance for three pediatric patience with varying physiology. (a) Heart valve segmentations in the mid-systolic configuration used to obtain the target leaflet surfaces. (b) FINESSE agreement with the target leaflet surface. Note larger errors are typically found in the commissure regions due to their omission from FINESSE to improve stability (see also Figure~\ref{Figure:ValveRegions}). (c) Assessment of BLDM in the central leaflet belly regions highlighting excellent agreement ($<0.3$\,mm) for all three patients on a similar order or better than the smallest voxel dimension (MV: $0.339$\,mm, CAVC: $0.889$\,mm, HLHS: $0.277$\,mm).}
\label{Figure:InVivoEvaluation}
\end{figure}

\begin{figure}[h]
\includegraphics[width=0.95\textwidth]{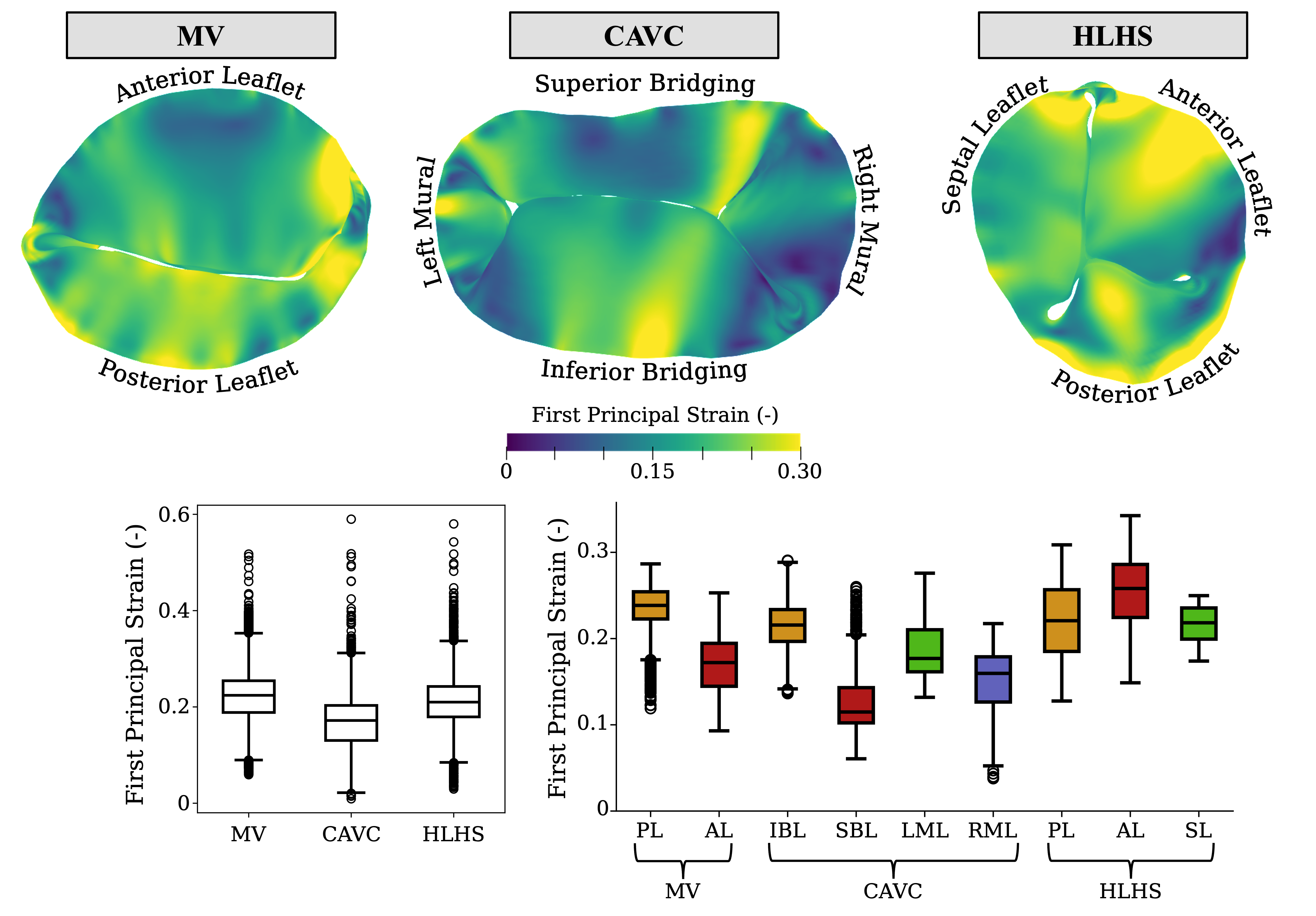}
\centering
\caption{\textit{In vivo} heart valve strains estimated using FINESSE for three pediatric patients with varying physiology. (a) Visualization of first principal strains for all three pediatric patients. (b) Comparison of the first principal strain distributions for the entire valve. (c) Comparison of the first principal strain distributions for the leaflet belly regions.}
\label{Figure:InVivoStrains}
\end{figure}

The FINESSE results can be further analyzed to understand differences in leaflet strains between the three cardiac physiologies (Figure\,\ref{Figure:InVivoStrains}(a)). All three valves have similar first principal strains when considering the entire valve geometry (MV: $0.224$\,$[0.188-0.254]$, CAVC: $0.172$\,$[0.130-0.203]$, HLHS: $0.210$\,$[0.179-0.242]$) (Figure\,\ref{Figure:InVivoStrains}(b)). However, there are unique comparisons that arise when isolating the central belly regions (Figure\,\ref{Figure:InVivoStrains}(c)). Starting with the patient with a CAVC, the inferior bridging leaflet (IBL) had larger first principal strains than the superior bridging leaflet (SBL). This parallels the results for the MV in which the posterior leaflet (similar anatomical position to IBL) had larger first principal strains than the anterior leaflet (similar anatomical position to the SBL). This suggests that the anatomical position may influence the leaflet strains, even when considering a complex congenital heart disease such as CAVC. Moreover, the TV in HLHS had large principal strains compared to both the MV and the CAVC despite being the youngest patient (2-days vs. 13-years). This underscores the challenging biomechanical environment for the TV in HLHS where  the right-side of the heart must undertake atypical systemic hemodynamics. Nevertheless, it is important to mention that these results are purely to demonstrate the \textit{in vivo} application of FINESSE, and our future studies will expand on these congenital heart disease populations. 

%%============%%
%% Discussion %%
%%============%%

\section{Discussion}
\label{Sec:Discussion}

\subsection{Overall Findings}

This study presents an open-source method for estimating \textit{in vivo} heart valve biomechanics using Finite Element Simulations with Shape Enforcement (FINESSE). We used a two-step method in which: (i) the systolic valve configuration is predicted using traditional forward FE simulations (e.g., \cite{wu2022, wu2023}) and (ii) a surface-to-surface registration algorithm is applied to enforce agreement between the FE simulation and the target systolic valve configuration. We have intentionally implemented this approach in FEBio, a NIH-supported open-source framework to ensure continued access, development, and refinement by the valve modeling community.

\subsubsection{Evaluation of FINESSE using Synthetic Data}
FINESSE was evaluated using three synthetic test cases ranging in complexity from a simple pressurized membrane to a stereotypical MV geometry. Our results demonstrated that FINESSE can provide excellent agreement with the target synthetic surface and synthetic tissue strains. The accuracy of FINESSE improved with a larger user-defined penalty, better agreement between the initial FE prediction and the target surface, and greater surface smoothness. Interestingly, the leaflet material model had minimal influence on the results, but softer materials tended to provide better agreement with the target surfaces. We further showed it is possible to compensate for the lack of image-derived chordae structures by either directly prescribing free edge motion or applying a CEF to the free edge nodes. Prescribed free edge motion better recaptures the known synthetic strain distributions, yet both methods have excellent agreement with the target surfaces and recapture median first principal strain. The final portion of the synthetic results demonstrated that FINESSE stability and performance improved when the commissures are not considered, and that the inability to precisely define the valve commissure and free edge geometries will not strongly impact FINESSE accuracy. 

\subsubsection{Application of FINESSE to \textit{In Vivo} Data}
In addition to verifying our approach in synthetic data, we applied FINESSE to 3DE \textit{in vivo} of three pediatric patients varying anatomy and physiology Our results showed that FINESSE successfully enforced the target \textit{in vivo} valve configuration for all three patients with error similar to the smallest voxel dimension. As expected, there were larger errors in the commissures due to omission from FINESSE to improve simulation convergence (see Section~\ref{sec:ResultsCardiacImagingChallenges}. All three valves had similar first principal strains with more pronounced differences found in the belly regions. For instance, we noted the IBL and SBL of the CAVC had similar first principal strains to the posterior and anterior leaflets of the MV, respectively, which may suggest anatomical leaflet position plays a role in \textit{in vivo} leaflet mechanics. Furthermore, the TV leaflets of the 2-day old patient with single ventricle physiology (e.g. HLHS) had larger principal strains than the MV and CAVC of 13-year old patients. This may underscore the unique challenges that the TV faces in HLHS, where the right-side of the heart must undertake atypical systemic hemodynamics. These findings demonstrate the need for further investigation of valve biomechanics in patients with single ventricle physiology and other populations with congenital heart disease.

\subsection{Comparisons with Existing Methods for Estimating \textit{In Vivo} Tissue Biomechanical Properties}

Estimating \textit{in vivo} tissue properties has received considerable attention for many soft tissues. Generally, these methods seek to use images of \textit{in vivo} tissue structures to optimize tissue properties (``Inverse Methods") or to inform/enforce tissue behavior (``Registration Methods"). There are other methods, such as elastography, that exploit wave propagation through materials to estimate tissue stiffness but these methods are outside the scope of the present work. Here, we provide a brief comparison of FINESSE to each class of methods with a focus on heart valve biomechanics where possible. 

\subsubsection{Inverse Methods}
The traditional approach for inverse methods is known as inverse finite element analysis (iFEA). In iFEA, forward FE simulation(s) are compared to the target \textit{in vivo} valve configuration, and constitutive model parameters are iteratively adjusted to improve agreement. Maniatty \textit{et al}.~\cite{maniatty1989} provide an excellent overview of early iFEA efforts for traditional solid mechanics and heat transfer problems. iFEA later emerged for soft tissue bimoechanics problems using tissue aspiration~\cite{kauer2002}, radiopaque markers placed in large animal models~\cite{krishnamurthy2008}, and bovine cornea tissue under pressure inflation~\cite{nguyen2011}, among others~\cite{erdemir2006, wittek2013}. This has allowed investigators to use \textit{in vivo} imaging to further elucidate differences in diseased tissues~\cite{narayanan2021, latorre2023, akyildiz2016}. In the heart valve biomechanics field, this has been done to estimate the \textit{in vivo} leaflet stresses for potential biomaterial design~\cite{lee2014}, develop patient-specific models~\cite{wang2013}, estimate patient-specific leaflet properties in congenital heart disease~\cite{ross2024}, among others~\cite{abbasi2016, aggarwal2016inverse, khang2023}. However, application of iFEA to heart valves is challenged by long optimization times and non-unique solutions that stem from valve functional complexity (self-contact, large deformations, buckling/bending), which has unfortunately precluded its widespread use.

Advances in machine learning techniques and high-performance computing have extended into estimating \textit{in vivo} tissue properties. A recent review article by Arzani \textit{et al}.~\cite{arzani2022} discusses the benefits of these approaches as well as challenges associated with with relying solely on machine learning for \textit{in vivo} assessment.  They recommended augmenting traditional engineering approaches (e.g., FE simulations) with machine learning methods as recently demonstrated for parameter estimation from experimental methods~\cite{kakaletsis2023, you2022} and soft tissue constitutive modeling~\cite{linka2023, he2021, holzapfel2021}. However, there is growing interest in the application of physics-augmented machine learning approaches to perform \textit{in vivo }parameter estimation of cardiac tissues~\cite{wu2024PINN}.

\subsubsection{Registration Methods}
A classical approach in registration methods involves using deformable templates of the soft tissue to estimate  \textit{in vivo} tissue deformations (e.g.,~\cite{mansi2011}). In principle, the resulting deformation field can be used to determine the tissue strain's without \textit{a priori} knowledge of the material properties. One approach, termed ``hyperelastic warping"~\cite{rabbitt1995, weiss1998}, uses the tissue material behavior for regularization to provide more realistic estimations of the deformation field. Furthermore, if the precise material properties are known, the tissue stress can then be directly estimated, providing rich biomechanical information. This method has been implemented into FEBio (see Figure 1 of \cite{maas2017}) and used successfully for the left ventricle~\cite{phatak2009, veress2005}, ligaments~\cite{phatak2007}, and coronary arteries~\cite{veress2002}. However, these structures are relatively simple compared to the heart valve leaflets, owing to valve-specific challenges: (i) large motions of the leaflets throughout the cardiac cycle and (ii) vanishing leaflet boundaries when against the surrounding cardiac tissue. While challenging, there have been recent efforts to overcome these challenges by augmenting a neural ordinary differential equation-based image registration framework with FE simulation synthetic information~\cite{wu2024}. 

A second approach is a methodology by which forward FE simulations are ``shape enforced" to match the valve structure derived from segmentations of \textit{in vivo} images. This approach circumvents the time-consuming, iterative nature of \textit{Inverse Methods} while ensuring realistic tissue deformations due to the use of FE simulations. Rego \textit{et al.} \cite{rego2018} were the first to pursue this method for heart valves using the commercial Abaqus solver (Dassault Syst\`emes Simulia Corp., Johnson, Rhode Island) with an in-house subroutine. They successfully demonstrated that shape enforcement could be used to estimate heart valve leaflet strains, and later used their approach to understand differences in strain patterns for patients with myocardial infarction \cite{narang2021, simonian2023, liu2023}. While this development was a significant step forward for the heart valve biomechanics community, its widespread use and clinical translation is limited by the use of a commercial software and in-house subroutine. For these reasons, we intentionally developed our FE simulation tools (including the present work) in the open-source FEBio software~\cite{wu2022, wu2023}. We believe this open-source philosophy will enable continuous refinement of heart valve FE simulation tools and the development of  seemless interfaces with open-source image analysis software such as Slicer~\cite{fedorov2012} and SlicerHeart~\cite{lasso2022}.

\subsection{Limitations and Future Improvements}
\label{sec:FutureExtensions}
This study is not without limitations. First, we have demonstrated the efficacy of FEBio FINESSE using synthetic FE data, but it is necessary to further validate FINESSE predictions using real experimental data. A prior work using a similar method showed that shape enforcement provides good estimations of \textit{in vivo} tissue strains derived using sonocrystals in sheep~\cite{rego2018}. We will pursue a similar experimental avenue in our forthcoming works. Second, our FE model fidelity is limited in regards to material anisotropy stemming from the underlying collagen fiber architectures (e.g.,~\cite{aggarwal2016inverse}). Although this may influence the orientation and magnitude of the in-plane principal strains, the inclusion of these factors should not impact the capability of FEBio FINESSE to match the target surface. Third, we are only considering one target (systolic) frame for our analysis. The heart valves are incredibly dynamic structures, so it would be worthwhile to enhance FEBio FINESSE to consider intermediate time points and fully recapitulate the cardiac cycle. Finally, limitations in 3D cardiac imaging prevent reconstruction of the chordae tendineae structures. This led us to consider two options for emulating the chordae function by means of directly prescribing free edge motion~\cite{ross2024} or prescribing a chordae emulating force~\cite{rego2018}. Although this enables FEBio FINESSE to recapture the closed valve configuration \textit{within one voxel of error}, we plan to explore generating ``functional" chordae structures using results from FEBio FINESSE. This will allow us to move towards standalone FE simulations that replicate \textit{in vivo} function and can be used to evaluate surgical interventions. 

\section{Concluding Remarks}
\label{Sec:Conclusion}

This study provided an open-source framework to perform finite element simulations with shape enforcement (FINESSE) in FEBio with applications to atrioventricular heart valves. Our method used a two-step process to first predict the mid-systolic valve configuration using traditional forward finite element simulations and then enforce surface-surface agreement with the target configuration via a surface-to-surface registration algorithm. We evaluated FINESSE performance using three synthetic test cases that ranged in complexity from a pressurized membrane to a stereotypical mitral valve geometry. Our results showed: (i) the \texttt{sliding-elastic} interface penalty and initial surface proximity strongly influence FINESSE performance, (ii) softer materials led to better predictions of the synthetic results, (iii) omitting the leaflet commissures from the shape enforcement improved FINESSE convergence and performance, and (iv) the chordae tendineae can be emulated by directly prescribing free edge motion or via a chordae emulating force. We concluded our investigation by using FINESSE to predict the \textit{in vivo} strains for three pediatric patients, two of which had complex congenital heart disease. FINESSE provided excellent agreement (approximately one voxel width) with the image-derived mid-systolic surfaces. The three valves had similar first principal strains when considering the entire valve geometry and more notable differences when only considering the leaflet belly regions. Of note, a 2-day old patient with HLHS generally had larger first principal strains than heart valves from two 13-year old patients (one with CAVC). We plan to expand on our comparisons to consider more pediatric patients with  congenital heart disease, and eventually link the image-derived strains with multi-scale tissue properties and long-term patient outcomes. Finally, it is important to emphasize that FINESSE is applicable to a wide-range of biological applications (e.g., arteries, ventricles, or tendons/ligaments), and we hope our open-source framework enables others to estimate \textit{in vivo} tissue strains for their topic of interest.

\backmatter

\bmhead{Acknowledgments}
This work was supported by the Additional Ventures Single Ventricle Research Fund, the Topolewski Pediatric Valve Center at CHOP, the Topolewski Endowed Chair, NIH 1R01HL153166, 2R01GM083925 (SAM, JAW), and T32 HL007915 (DWL). We would also like to thank the CHOP Research Institute for providing high performance computing resources.

\newpage
\begin{appendices}

\section{Bi-Directional Local Distance Measure}
\label{Sec:AppendixA}

This study used the bi-directional local distance measure (BLDM) to assess FINESSE accuracy (Algorithm~\ref{alg:BLDM}). The BLDM was defined as the maximum of the ``forward minimum distance" ($d_{\textrm{F-Min},i}$) and ``backward maximum distance" ($d_{\textrm{B-Max},i}$) for any node $p_{\textrm{FE},i}$ of the FINESSE-predicted surface ($\mathcal{S}_{\textrm{FE}}$). First, $d_{\textrm{F-Min},i}$ was determined as Euclidean distance between $p_{\textrm{FE},i}$ and the closest node $p_{\textrm{Target},j}$ on the target surface ($\mathcal{S}_{\textrm{Target}}$). Second, $d_{\textrm{B-Max},i}$ was determined in a two-step fashion: (i) find all $p_{\textrm{Target},j}$ with $p_{\textrm{FE},i}$ as the closest point on ($\mathcal{S}_{\textrm{FE}}$), (ii) determine the maximum Euclidean distance between all $p_{\textrm{Target},j}$ and $p_{\textrm{FE},i}$. The BLDM for the $i$-th node of the FINESSE-predicted surface was then taken to be the maximum value of $d_{\textrm{F-Min},i}$ and $d_{\textrm{B-Max},i}$.

The BLDM is a nodal error calculation that can result in non-smooth error distributions when nodal density of the two surfaces differs substantially. We therefore added an additional consideration when computing $d_{\textrm{F-Min},i}$ to leverage the element construction of $\mathcal{S}_{\textrm{Target}}$ to produce smoother BLDM maps. Rather than using the Euclidean distance between $p_{\textrm{FE},i}$ and the closest node on the target surface ($\mathcal{S}_{\textrm{Target}}$), we projected $p_{\textrm{FE},i}$ onto the closest \textit{element} of the target surface (determined via the centroid location). This resulted in smoother plots of the BLDM (e.g., Figure~\ref{Figure:InVivoStrains}) compared to prior works leveraging the BLDM for surface-surface error quantification (e.g., Figure 7 of \cite{ross2024}).

\begin{algorithm}
\caption{Modified bi-directional local distance measure calculation.}
\label{alg:BLDM}
\begin{algorithmic}[1]
\State{$p_{\textrm{Target}}$: $\mathcal{S}_{\textrm{Target}}$ nodes}
\State{$p_{\textrm{FE}}$: $\mathcal{S}_{\textrm{FE}}$ nodes}
\State{$e_{\textrm{Target}}$: $\mathcal{S}_{\textrm{Target}}$ elements}
\\
\Procedure{BLDM}{$p_{\textrm{Target}}$, $p_{\textrm{FE}}$, $e_{\textrm{Target}}$}
\For{$j$ in $p_{\textrm{Target}}$}\Comment{Pre-populate IDs for $d_{\textrm{B-Max}}$}
    \For{$i$ in $p_{\textrm{FE}}$}
        \State{$d[i] \gets \left\lVert{p_{\textrm{Target}}[j] - p_{\textrm{FE}}[i]}\right\rVert$}
    \EndFor
    \State $\textrm{ID}_{\textrm{back}}[j] \gets \textrm{argmax}(d)$
    \State $d_{\textrm{back}}[j] \gets \textrm{max}(d)$
\EndFor
\\

\For{$i$ in $p_{\textrm{FE}}$}\Comment{Determine BLDM for each $p_{\textrm{FE}}$}
    \For{$j$ in $p_{\textrm{Target}}$}\Comment{Find ID for $d_{\textrm{F-Min}}$}
        \State{$d[j] \gets \left\lVert{p_{\textrm{FE}}[i] - p_{\textrm{Target}}[j]}\right\rVert$}
    \EndFor
    \State $\textrm{ID}_{\textrm{F-Min}} \gets \textrm{argmin}(d)$
    \\
    \State{$e_{\textrm{connect}} \gets e_{\textrm{Target}}$ containing $\textrm{ID}_{\textrm{F-Min}}$}\Comment{Find closest element with $\textrm{ID}_{\textrm{F-Min}}$}
    \For{$k$ in $e_{\textrm{connect}}$}
        \State{$c[k] \gets $ centroid of $e_{\textrm{connect}}[k]$}
        \State{$d[k] \gets \left\lVert{c[k] - p_{\textrm{FE}}[i]}\right\rVert$}
    \EndFor
    \State{$\textrm{ID}_{\textrm{ele}} \gets \textrm{argmin}(d)$}
    \State{$n_{\textrm{ele}} \gets \textrm{normal vector of }e_{\textrm{connect}}[\textrm{ID}_{\textrm{ele}}]$}
    \State{$c_{\textrm{ele}} \gets c[\textrm{ID}_{\textrm{ele}}]$}
    \\
    \State{$d_{\textrm{F-Min}}[i] \gets \left|(p_{\textrm{FE}}[i] - c_{\textrm{ele}}) \cdot n_{\textrm{ele}}\right|$}
    \State{$d_{\textrm{B-Max}}[i] \gets \textrm{argmin}(d_{\textrm{back}}[\textrm{ID}_{\textrm{B-Max}} == p_{\textrm{FE},i}])$}
    \State{$\textrm{BLDM}[i] \gets \textrm{argmin}(\{d_{\textrm{F-Min}}[i], d_{\textrm{B-Max}}[i]\})$}
\EndFor
\State{\textbf{return} BLDM}
\EndProcedure
\end{algorithmic}
\end{algorithm}
\newpage
\section{Generating FINESSE Surfaces from 3DE Images}
\label{Sec:AppendixB}

The FE mesh and FINESSE target surfaces were created using custom SlicerHeart modules in the open-source 3D Slicer software. The following details the step-wise process with the assumption that the patient-specific segmentation has been created. Details for the segmentation process can be found in Supplemental Video 3 of \cite{lasso2022}.

\begin{enumerate}
    \item{Segmentation Pre-Processing}
    \begin{description}
        \item[Note:]{This step is repeated for both the open (mid-diastole) and closed (mid-systole) valve segmentations. }
    \end{description}
    \begin{itemize}
        \item{Navigate to the \texttt{Data} module}
        \item{Export each leaflet segmentation to a model via right click $\rightarrow$ ``Export visible segments to models"}
        \item{Navigate to the \texttt{Surface Toolbox} module}
        \item{Decimate each leaflet model by $80\%$, resulting in approximately $3000$-$4000$ elements}
        \item{Smooth each leaflet model}
        \begin{itemize}
            \item{Method: Taubin}
            \item{Iterations: $30$}
            \item{Pass band: $0.10$}
            \item{Boundary smoothing: Disabled}
        \end{itemize}
    \end{itemize}
    \item{Medial Surface Construction~\cite{herz2024}}
    \begin{description}
        \item[Note:]{This step is repeated for both the open (mid-diastole) and closed (mid-systole) valve leaflet models.}
    \end{description}
    \begin{itemize}
        \item{Navigate to the \texttt{Synthetic Skeleton} module}
        \begin{itemize}
            \item{Edge Criterion: $4$}
            \item{Pruning Factor: $0.70$}
            \item{Max Conn. Component: $0$}
            \item{Tolerance: $0.00$}
        \end{itemize}
        \begin{description}
            \item[Note:]{It is possible this will result in a non-triangular mesh and the user may need to test different combinations of the \texttt{Synthetic Skeleton} I/O.} 
        \end{description}
        \item{Navigate to the \texttt{Surface Toolbox} module}
        \item{Compute surface normals with ``Flip normals" selected}
        \item{Uniform remesh}
        \begin{itemize}
            \item{Number of points: $20.0$k}
            \item{Subdivide: $2$}
        \end{itemize}
        \item{Smooth the resulting surface}
        \begin{itemize}
            \item{Method: Laplace}
            \item{Iterations: $50$}
            \item{Relaxation: $0.1$}
            \item{Boundary Smoothing: Disabled}
        \end{itemize}
        \begin{description}
            \item[Note:]{Larger values of \texttt{Iterations} and \texttt{Relaxation} may substantially smooth the target surfaces. Our experiences suggests this will prevent leaflet coaptation, and may flatten the leaflets. The provided values are suggestions, and the user should check the quality of the smoothed surface.} 
        \end{description}
        \item{Decimate the smoothed surface by $90\%$}
    \end{itemize}
    \item{Valve Landmark Definition}
    \begin{description}
        \item[Note:]{This step is repeated for both the open (mid-diastole) and closed (mid-systole) valve leaflet models.}
    \end{description}
    \begin{itemize}
        \item{Navigate to the \texttt{Markups} module}
        \item{Create a \texttt{Closed Curve} for the valve annulus and free edge.}
        \item{Create a \texttt{Point List} for the commissures along the annulus and free edge closed curves.}
        \begin{description}
            \item[Note:]{These curves and point lists should be defined in the same direction along the annulus.} 
            \item[Note:]{Have the closed curves be extensions of the medial surfaces that intersect with the valve segmentation.}
        \end{description}
    \end{itemize}
\end{enumerate}

The remainder of the preparation occurred within custom code that is available upon request. First, the resulting annulus and free edge curves in mid-diastole were lofted to create the leaflet surface, which was then meshed using FEBio or any preferred meshing software. Additional constrains such that the lofted surface aligns with image-derived medial leaflet surface can be incorporated if desired. Second, the annulus curves in mid-diastole and mid-systole were used to determine patient-specific annular displacements for the FE simulation. Third, the acquired medial leaflet surface in mid-systole were imported into FEBio as a fixed rigid body and considered as the target surface in the \texttt{sliding-elastic} interface.

\end{appendices}

\newpage
% \bibliography{bibliography}

\end{document}